\documentstyle[epsfig]{aipproc}

\begin{document}
\title{Fast Variations of Gamma-Ray Emission in Blazars }

\author{Stefan J. Wagner$^*$, Corinna von Montigny, and Martin Herter}
\address{$^*$swagner @ lsw.uni-heidelberg.de \\ 
Landessternwarte, K\"onigstuhl, 69117 Heidelberg, Germany}

\lefthead{S.J. Wagner et al.}
\righthead{Fast Variations of Gamma-Ray Emission in Blazars}
\maketitle

\begin{abstract}
The largest group of sources identified by EGRET are Blazars. This
sub-class of AGN is well known to vary in flux in all energy bands on
time-scales ranging from a few minutes (in the optical and X-ray
bands) up to decades (radio and optical regimes). In addition to
variations of the gamma-ray flux between different viewing periods,
the brightest of these sources showed a few remarkable gamma-ray
flares on time-scales of about one day, confirming the extension
of the ``Intraday-Variability (IDV)'' phenomenon into the GeV range. 
We present first results of a systematic approach to study fast
variability with EGRET data. This statistical approach confirms
the existence of IDV even during epochs when no strong flares
are detected. This provides additional constraints on the site of
the gamma-ray emission and allows cross-correlation analyses with
light curves obtained at other frequencies even during periods
of low flux.\\
We also find that some stronger sources have fluxes 
systematically above threshold even during quiescent states.
Despite the low count rates this allows explicit comparisons
of flare amplitudes with other energy bands.
\end{abstract}

\subsection*{Introduction} 
Most of the identified sources in the energy
band studied with EGRET are Blazars. One of the defining criteria of
this class is pronounced variability in the optical and radio bands
on time-scales of weeks to months \cite{stein}. Later it was
realized that these sources often show strong and variable X-ray
emission as well. During the last decade it became clear that
variations can be traced down to much shorter time-scales in all
energy bands. Rapid changes ``IDV'' \cite{ww95}
can now be traced down to the shortest time-scales which can be probed
in all of the different wavelength regimes. In the radio domain
variations have been found to occur within less than two hours
\cite{qui89}, \cite{kra97}, \cite{lkc97}, \cite{wsoc97}, in the
optical and near-IR regimes variations on time-scales of hours
\cite{wea96}, \cite{mil95}, \cite{tak96} down to a few minutes
\cite{wsoc97} have been reported. In the X-ray regime variations have been
reported to be even faster \cite{fei82}, \cite{rem91}, \cite{wea96}.

\begin{figure}[b!] 
\centerline{\hspace*{3mm}
\epsfig{file=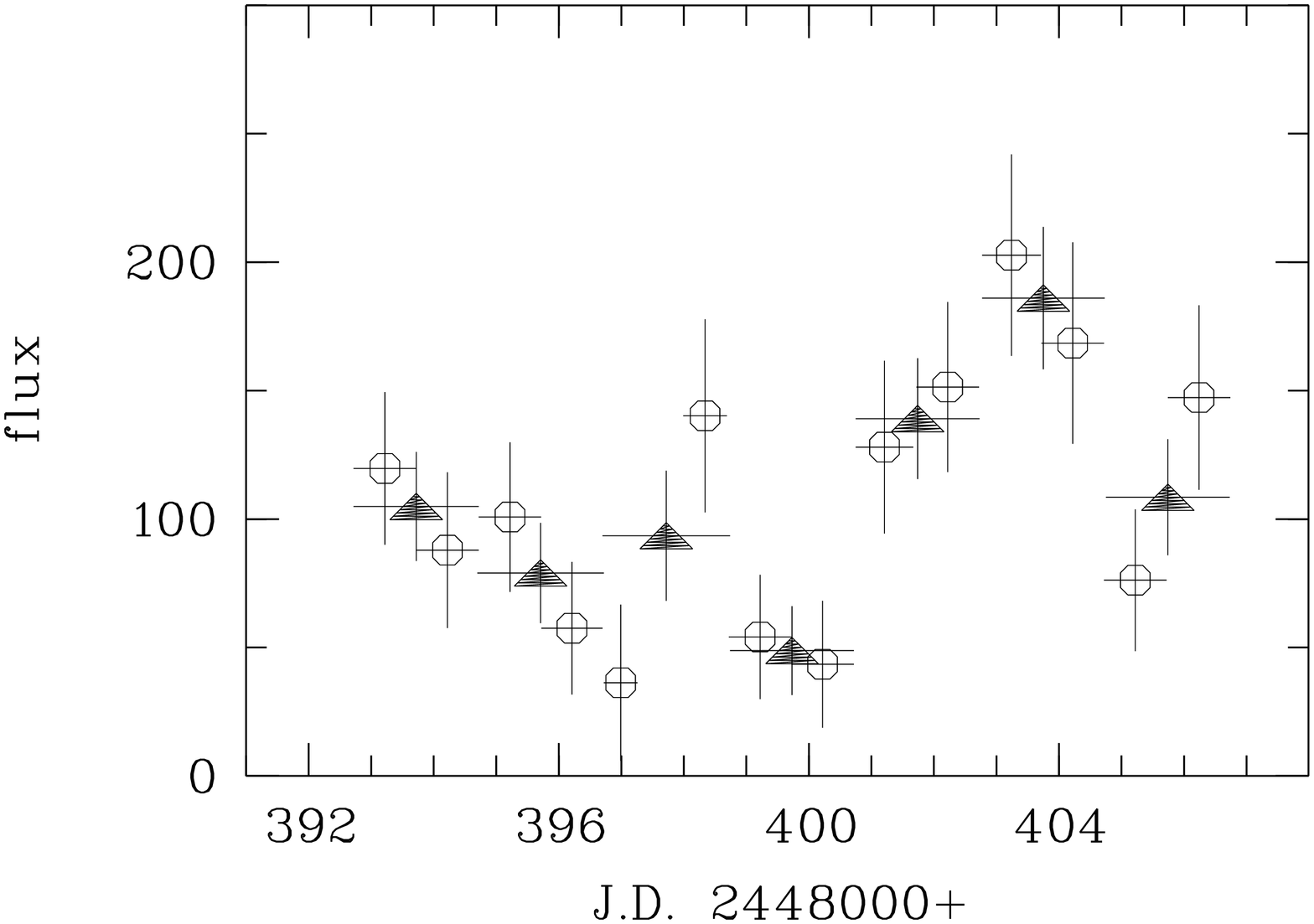,angle=-90,height=1.55in,width=2.8in}
\epsfig{file=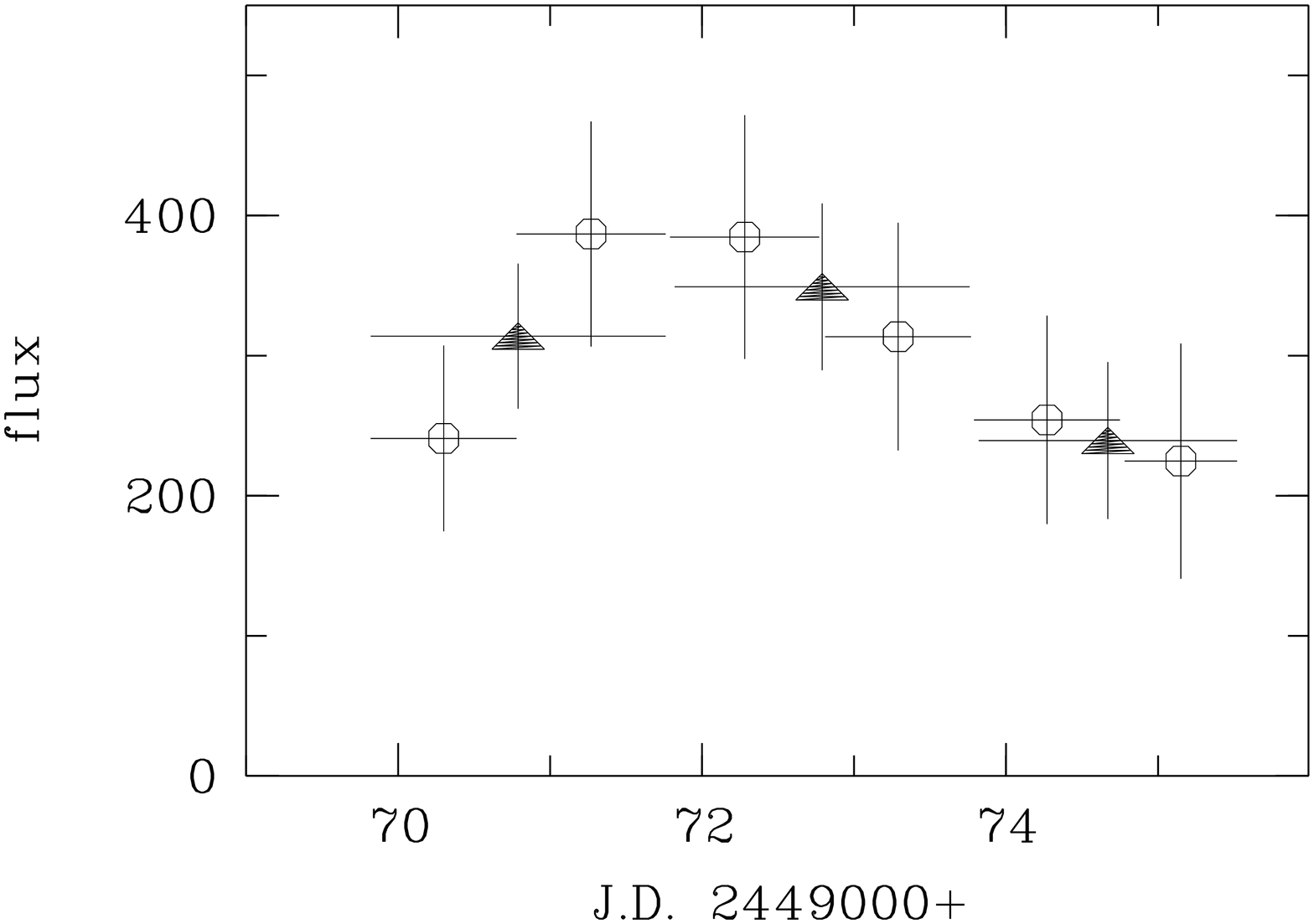,angle=-90,height=1.55in,width=2.8in}
}
\vspace{10pt}
\caption{Light curves of PKS 0528+134 with a binning of one (open circles) 
and two days (full triangles) during two different viewing periods (flux is
given in 10$^{-8}$ $\gamma$ cm$^{-2}$ s$^{-1}$). While
$\chi^2$ tests indicate significant variations in the left panel (not in 
the right one), a two day binning is too poor to reveal the shortest 
time-scales.}
\label{fig1}
\end{figure}

Blazars also turned out to be the dominant class of sources in the GeV regime.
It was soon realized that they vary from one EGRET viewing period to another
(von Montigny, et al., this meeting). If observed during a particularly
bright state, individual objects were shown to vary on shorter time-scales
as well (e.g. \cite{knif93}). With typical recorded flux-densities of about
a photon per hour it is impossible to probe the regime of very fast variations
($<$ hours) which can be tackled in other energy bands. Statistical approaches 
enable investigations of the fastest variation that {\it can} be studied.
They also allow comparisons throughout the entire range of time-scales with
those in other regimes of the electromagnetic spectrum.

\subsection*{Rapid Gamma-ray Flares}
Already one of the first EGRET pointings towards a Blazar found the source
3C\,279 in a state of enhanced activity. Detailed analysis revealed
variations on time-scales of a few days with high significance \cite{knif93}, 
\cite{hea96}. 
Comparable variations were found in other sources, including,
e.g., PKS 0528+134 \cite{hun93}, PKS 1406-076 \cite{wea95}, PKS 
1633+382 \cite{matt93}, and PKS 2251+158 \cite{hea93}). Hunter et al. 
\cite{hun93} illustrated that the variability of PKS 0528+134 was 
persistent over at least eight weeks with several maxima rather than 
a singular, well defined peak. This is comparable to the variations of 
the synchrotron emission of Blazars in general, as seen e.g. in the 
optical regime \cite{ww95}. In an attempt to constrain the shortest 
time-scales, Mattox et al. \cite{matt96} analysed the brightest flare of 
any gamma-emitting Blazar, PKS~1633-29, by binning the light curve with 
a variable width in time such that the individual bins had comparable 
errors. This revealed time-scales as short as about 6 hours in the 
observers frame.

In order to study the statistical properties of fast variations we
analysed the EGRET observations of a larger set of Blazars on
time-scales of about one day. All observations of a selected source
available from the EGRET archive were taken and binned into one-day
maps. This binning was chosen to remain sensitive to the fastest
changes seen in EGRET data so far. Rebinning to longer time-scales 
could be performed without any significant
degradation in accuracy. Examples of light curves binned into
intervals of one and two days, respectively, are shown in Figure
\ref{fig1}.

\begin{figure}[t!] 
\centerline{\epsfig{file=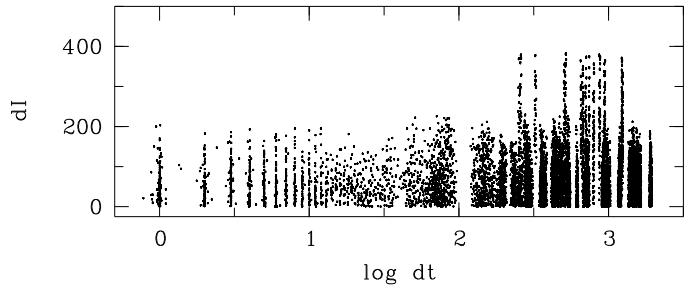,angle=-90,height=1.55in,width=4.0in}}
\vspace{10pt}
\caption{Differences in intensity of PKS 0528+134 between any two epochs as 
a function of the difference in time log(dt/1 day) between the measurements.}
\label{fig3}
\end{figure}

The entire data set of every well-observed source can be studied in 
over 100 one-day maps.
This is a sufficiently large data base to carry out
statistical studies, comparable to the structure function analysis used
in temporal studies at other wave-bands. A similar method has been used by
\cite{wea96} to study IDV in 0716+714 in the radio and
optical bands. 

\subsection*{Statistical Analyses}

\begin{figure}[t!] 
\centerline{\hspace*{3mm}
\epsfig{file=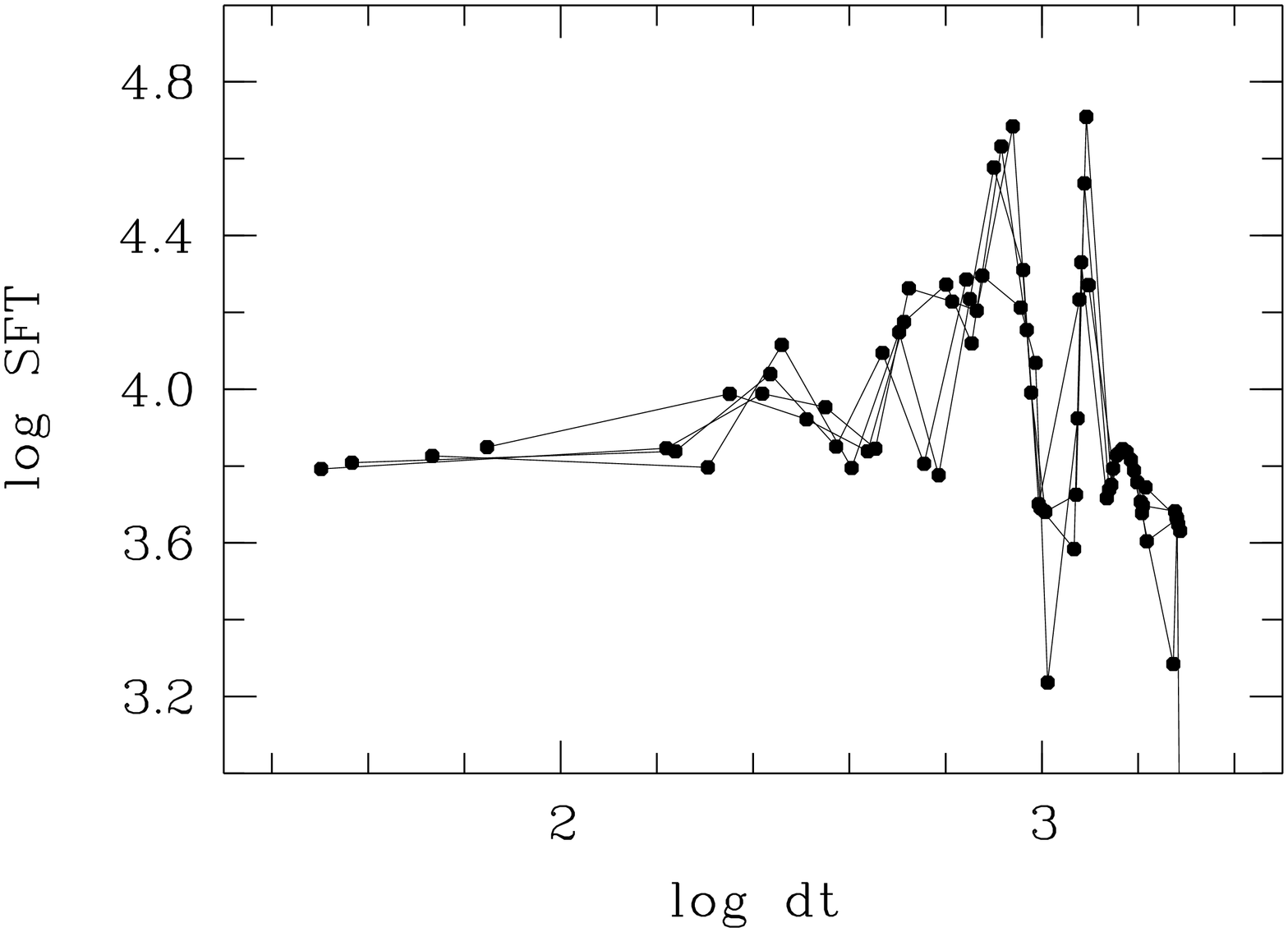,angle=-90,height=1.5in,width=1.85in}
\epsfig{file=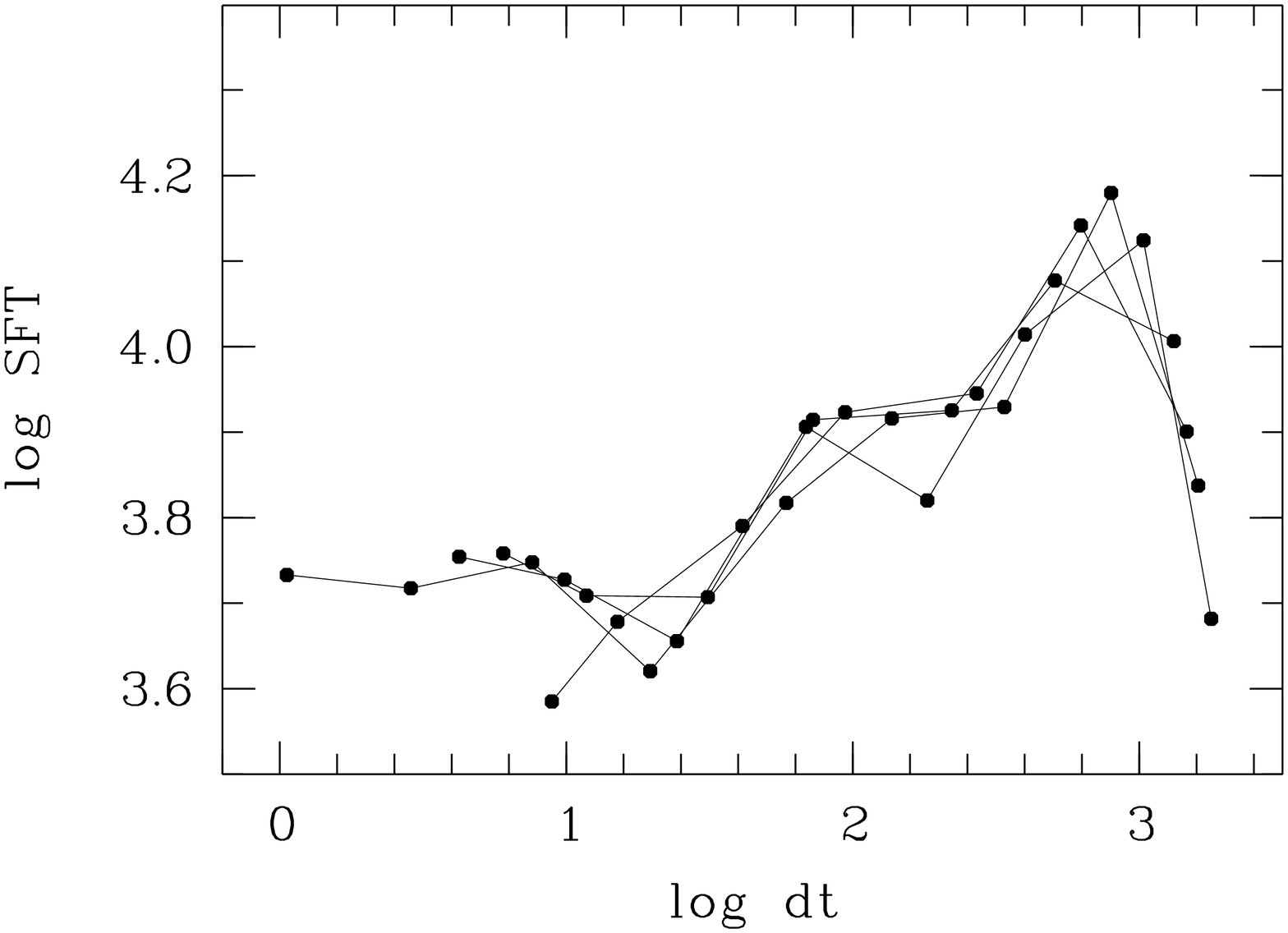,angle=-90,height=1.5in,width=1.85in}
\epsfig{file=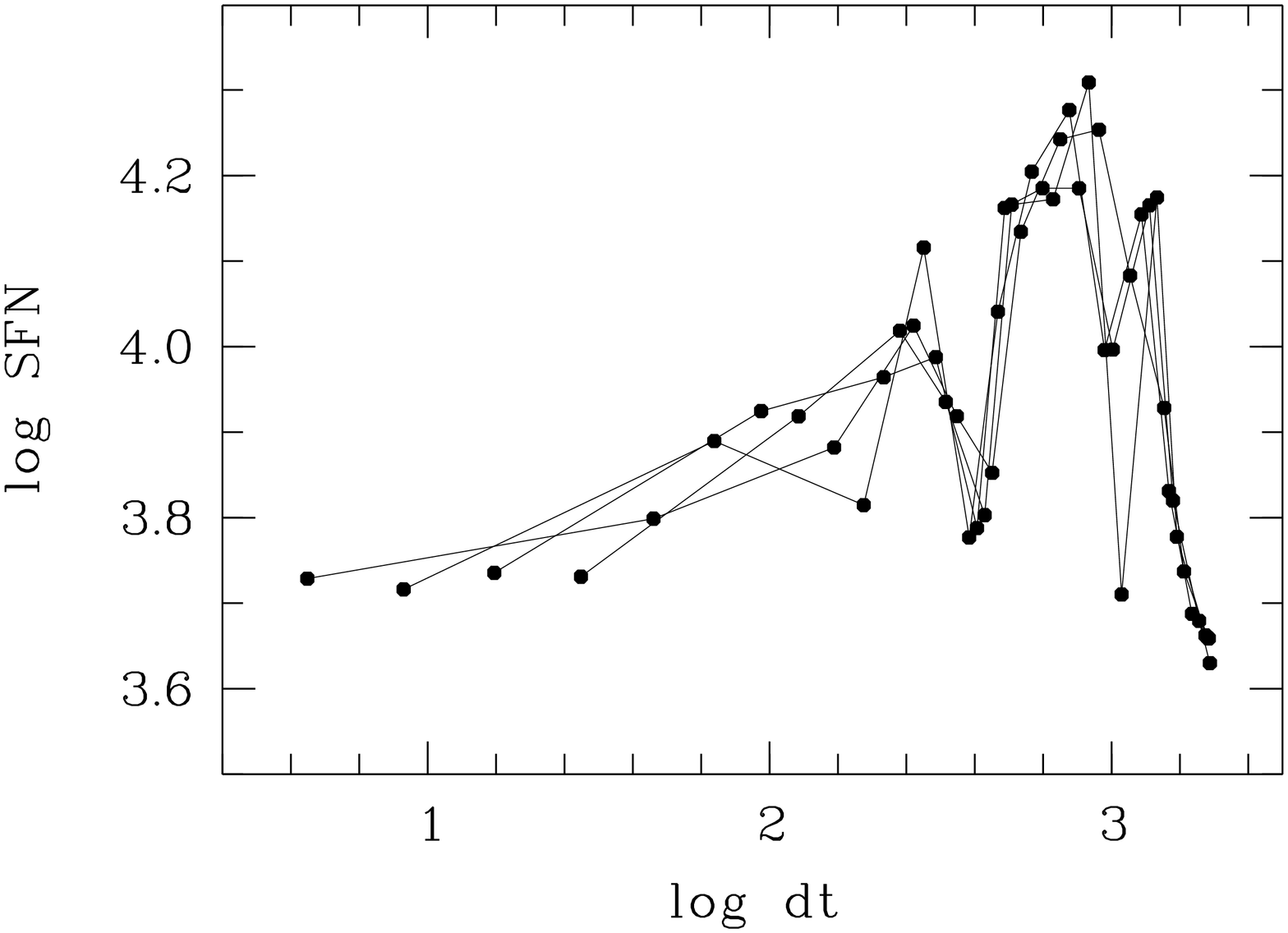,angle=-90,height=1.5in,width=1.85in}
}
\vspace{10pt}
\caption{Structure functions of PKS 0528+134 derived from bins of dt = 100
days (left panel, phase offsets of 25 days), log(dt/1 day) = 0.4 (central
panel, phase offsets of log(dt/1 day) = 0.1), and n = 800 (right panel,
phase offset of dn = 200).}
\label{fig4}
\end{figure}

All light curves of a source, binned in one-day intervals (as shown in
Figure \ref{fig1}), are combined and used to create structure
functions. In order to determine characteristic temporal changes in
flux, every combination of two measurements I(t$_i$), I(t$_j$) is used
to derive one measurement dI = I(t$_i$)-I(t$_j$) at dt =
(t$_i$-t$_j$). The entire data set of PKS 0528+134 is given in Figure
\ref{fig3}. The structure function as defined by \cite{wea96} was
derived from the set of dI(dt) by averaging a) equal bins in dt, b)
equal bins in log(dt), and c) bins of equal numbers of measurement
points dI(dt). With the small number of measurements, the uncertainty
introduced by choosing a particular binning and phase may be
significant.  Figure \ref{fig4} illustrates the resulting structure
functions, derived by using the three different ways of binning. Each
panel gives four curves, representing shifts of the phase by 25 \% of
the width of the bins.

\begin{figure}[b!] 
\centerline{\hspace*{-0.2in}
\epsfig{file=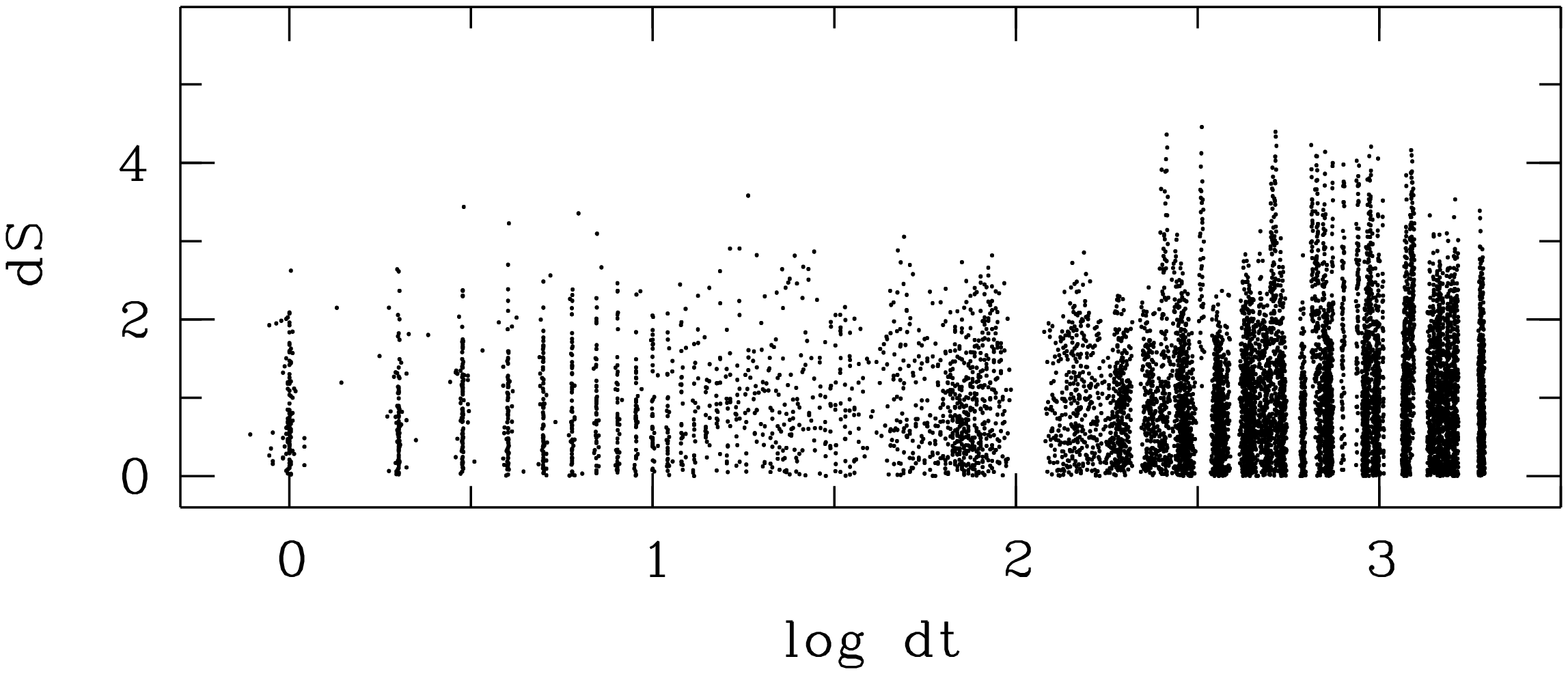,angle=-90,height=1.48in,width=3.5in}
\epsfig{file=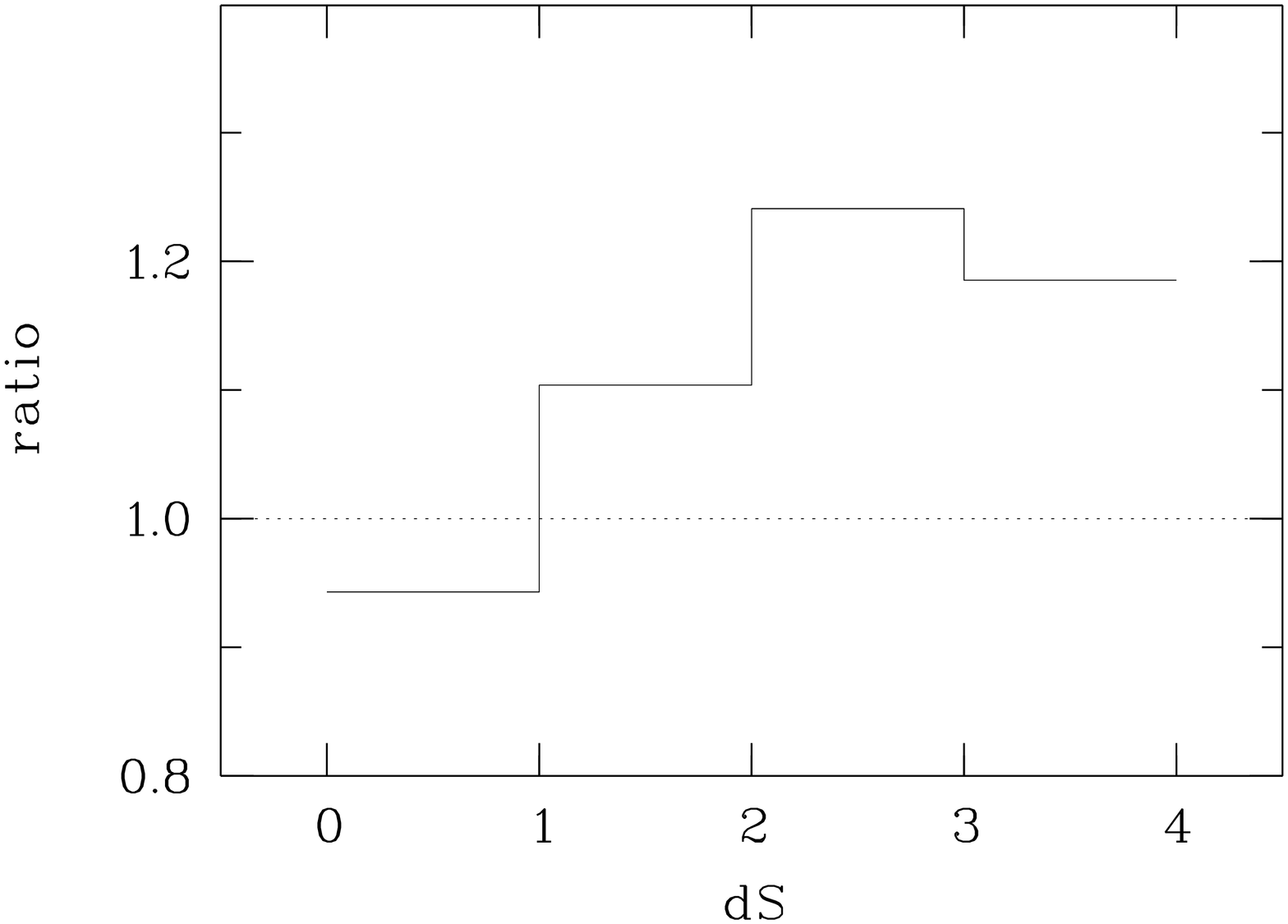,angle=-90,height=1.45in,width=2.3in}}
\vspace{10pt}
\caption{{\it Left}: dS(dt) distribution - differences in intensity between any
two points, normalised to their significance. {\it Right}: Ratio of actual
dS(dt) (for dt $\sim$ 1 day) for PKS 0528+134 to those of the normal 
distribution for a constant source. There is a clear excess of differences
at the 2-3 Sigma level, indicating variability on a statistical level.}
\label{fig5}
\end{figure}

Since the exposure of different one-day maps changed dramatically, the
errors on the dI(dt) varied significantly. This is taken into account
in the computation of the structure functions by individual weights but
is not illustrated with error bars in Figure \ref{fig3} for clarity.
Instead we present another version of Figure \ref{fig3} 
where the dI(dt) are given in terms of their significance dS(dt) = 
dI(dt)/$\sigma_{dI(dt)}$ (Figure \ref{fig5}, left panel). The right panel 
compares the histogram of dS(dt) for dt $\sim$ 1 day with the normal
distribution expected for a steady source with random errors, which clearly 
illustrates the excess at large dS.

Interpreting the results shown in Figure \ref{fig5} directly as evidence
for the statistical occurrence of variability relies on an accurate
determination and propagation of errors. We consistently chose conservative
errors.

We tested different approaches
by either fixing the positions of the sources or leaving them as free 
parameters and by fixing the background or leaving it as free parameters.
Despite subtle differences, we arrived at consistent results. 

As an alternative approach we treated other sources within the same
pointing directions. If errors are not estimated correctly, one may still
regard the least variable source within any field of view as constant and
derive the actual spread in errors from the dispersion of those sources
with least variability. This will allow the identification of variable
sources even if there are unknown contributions to the total errors as long
as constant sources exist. Although gamma-bright pulsars may be considered
as being constant on time-scales much longer than their pulsation periods 
and much shorter than cooling time-scales, we found significant differences
between the structure functions of the Crab Pulsar and Geminga. These results
will be presented separately.

\subsection*{Results and Discussion} 

We computed structure functions of several gamma-bright Blazars using
light curves obtained from maps with a temporal binning of one day. IDV 
is detected in the EGRET data on a significant level even during
those observations when the sources are not exceptionally bright. Gamma-ray IDV
has a high duty cycle and is not confined to individual flares. This is
similar to the variability seen at lower photon energies, and permits
cross-correlation studies even if IDV remains at low amplitudes.
Comparing the fluxes of Blazars during their low states with those derived at
fixed but ``empty'' positions in the same one-day maps, we find a statistical
indication for PKS 0528+134 and 0716+714 to have a steady flux about 1 sigma
above the background. This is consistent with previous upper limits but
confirms that EGRET data are sensitive enough to reach even the lower
envelope of the range in flux densities.

\end{document}